# A Magnetically-Supported PDR in M17

E.W. Pellegrini & J.A. Baldwin

*Physics and Astronomy Department, Michigan State University, 3270 Biomedical Physical Sciences Building, East Lansing, MI 48824*

C.L. Brogan

*National Radio Astronomy Observatory, 520 Edgemont Rd, Charlottesville, VA 22903*

M.M. Hanson

*Department of Physics, The University of Cincinnati, Cincinnati, OH 45221-0011.*

N.P. Abel, G.J. Ferland, H.B. Nemala, G. Shaw & T.H. Troland

*Department of Physics and Astronomy, University of Kentucky, 177 Chemistry/Physics Building, Lexington, KY 40506*

baldwin@pa.msu.edu

## Abstract

The SW part of the bright Galactic H II region M17 contains an obscured ionization front that is most easily seen at infrared and radio wavelengths. This "SW bar" has received considerable attention because the ionization front is seen nearly edge-on, thus offering an excellent opportunity to study the way in which the gas changes from fully ionized to molecular as radiation from the ionizing stars penetrates into it. M17 also is one of the very few H II regions for which the magnetic field strength can be measured in the photodissociation region (the "PDR") that forms the interface between the ionized and molecular gas. Here we carefully model an observed line of sight through the gas cloud, including the $H^+$, $H^0$ (PDR) and molecular layers, in a fully self-consistent single calculation. An interesting aspect of the M17 SW bar is that the PDR is very extended. We show that the relatively strong magnetic field which is observed to be present inevitably leads to a very deep PDR, because the structure of the neutral and molecular gas is dominated by magnetic pressure, rather than by gas pressure as previously had been supposed. We also show that a wide variety of observed facts can be explained if a hydrostatic geometry prevails, in which the gas pressure from an inner x-ray hot bubble and the outward momentum of the stellar radiation field compresses the gas and its associated magnetic field in the PDR, as has already been shown to occur in the Orion Nebula. The magnetic field compression may also amplify the local cosmic ray density by a factor of 300. The pressure in the observed magnetic field just balances the outward forces, suggesting that the observed geometry is a natural consequence of the formation of a star cluster within a molecular cloud.

*Subject headings:* H II regions –ISM: atoms – ISM: magnetic fields – ISM: individual (M17)

# 1. Introduction

A major focus of modern astrophysics is to understand the various feedback mechanisms involved in the cycle of star formation, star death, and the chemical enrichment of galaxies. Star formation is a key step in this continuous chemical and also structural evolution of the universe, and understanding the processes involved in it is an important ongoing topic for the current generation of Great Observatories and large ground-based telescopes. This subject will truly move to the forefront when JWST, ALMA and 30m ground-based telescopes come on line in the next decades, with the huge strides forward that they will bring in our ability both to study very distant, highly redshifted galaxies, and to see deep into local star-forming regions. A long term goal of our work is to learn how to better interpret the spectra of giant starbursts seen at high redshifts by studying nearby examples which have different levels of complexity, and use them to calibrate the nature of their more luminous but also more distant and therefore unresolved cousins.

The Omega Nebula, M17, is a bright galactic H II region which represents an intermediate step in complexity between the Orion Nebula (which is mostly ionized by a single star) and the nearest true Giant HII Regions (such as NGC 3603 and 30 Doradus, each of which is ionized by hundreds of O stars). M17 is ionized by a compact (but heavily obscured) cluster with perhaps a dozen O stars. Figure 1 summarizes some of the extensive multi-wavelength mapping of its projected structure, as well as showing the slit position at which we obtained the new optical spectroscopy described below in §4. The grey-scale image in the figure shows the visible light, in which we see mainly a bright bar roughly at PA 130°, which we will refer to as the "NE bar". The heavy contours show 21 cm radio continuum due to thermal bremsstrahlung. The radio contours are a reddening-free map of the H II region, and show another bar structure to the SW, lying along PA ~ 150°, which we will call the "SW bar". Evidently the SW bar is an edge-on ionization front that is heavily obscured by dust mixed in with molecular and atomic gas along the line of sight. The figure also shows, as narrow contours, the CO emission from the molecular cloud that is adjacent to the ionized gas.

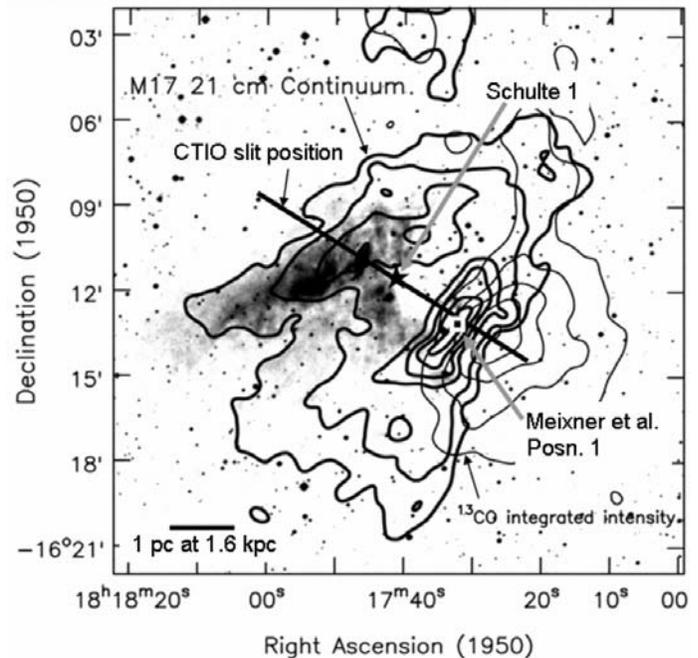

**Figure 1.** M17 in visible light (gray scale image), $^{13}$CO emission (narrow contours), and 21 cm radio continuum emission (heavy contours). The straight line shows the CTIO slit position. The star symbol indicates the star Schulte 1, which was used as a reference for positioning the spectrograph slit and which is referred to in the text. The small square marks Postion 1 from Meixner92, which is the line of sight that we analyze here. Figure adopted from BT01, and includes CO data from Wilson et al. 1999.

The goal of this paper is to determine the physical conditions in the SW bar region of M17 by comparing the predictions of a photoionization model with observations. We use the equilibrium photoionization code "Cloudy" (Ferland et al. 1998), which is unique in that it produces a fully self-consistent model including the $H^+$, $H^0$ (PDR) and $H_2$ regions, taking into account the effects of dust and the reaction networks for the most prominent molecular species (Abel et al. 2005) These calculations will then tell us the radiation, kinetic and magnetic pressure contributions in the PDR and molecular gas.

Our reason for selecting M17 from among other star-forming regions of similar size is that the edge-on ionization front permits a detailed study of ionization as a function of depth into the cloud. Another and even more important reason is that M17 one of the very few H II regions for which a magnetic field map has been made for the adjacent atomic gas (BT99; BT01), and has by far the simplest geometry of these cases. The magnetic field measurement offers an unusual opportunity to directly measure the equation of state, the relationship between temperature and sources of pressure within a PDR. Previous studies of H II regions usually have assumed that gas pressure dominates, so that if there is pressure balance between these two zones the atomic gas should be about 200 times denser than the $H^+$ (a factor of 100 from the temperature difference, and an additional factor of two from the recombination). At the other extreme, if the gas is completely magnetically and/or turbulently supported, the ionized and neutral gas would have the same hydrogen densities. These alternatives correspond to quite different cloud structures, with drastically different thicknesses for the PDR and a major impact on the physical conditions and emitted spectrum of the gas (see Abel & Ferland 2006). This present paper for the first time combines direct measures of the gas pressure in the $H^+$ layer and the magnetic pressure in the adjacent PDR, together with a full simulation of the physical processes in the gas, allowing a direct observational measurement of the equation of state.

## 2. The Geometry of M17

### 2.1 Overall Structure

M17 is a classic example of a blister H II region, in which the ionization front is moving into a molecular cloud which has recently formed the ionizing stars. Figure 2 shows the geometrical model that we will develop below and use throughout this paper. Because we see the ionization front in the SW bar nearly edge-on, M17 affords an excellent opportunity to study the interface between the fully ionized region nearer the star cluster and the PDR (the "photo-dissociation region"), the transition region where the gas becomes atomic and molecular. Thus M17 is a nearly perfect complement to Orion, where we see the same geometry nearly face on (Baldwin et al. 1991).

The distance to M17 has been debated for quite some time now. Chini et al, (1980) derived a distance of 2.2±0.2 kpc based on UBV photometry of 19 possible early type stars. Hanson, Howarth & Conti (1997) derived a distance of 1.3(+0.4/ -0.2) kpc based on published apparent and absolute magnitudes derived from K-band spectroscopy. Nielbock et al (2001) have deduced a distance of 1.6±0.3 kpc, which is the distance that is adopted throughout this paper. The decrease compared to the early observation of Chini et al results from a better determination of the extinction law and the elimination of the infrared excess objects. For a 1.6 kpc distance, each of the ionized bars is about 4 pc long and about 0.5 – 1 pc wide.

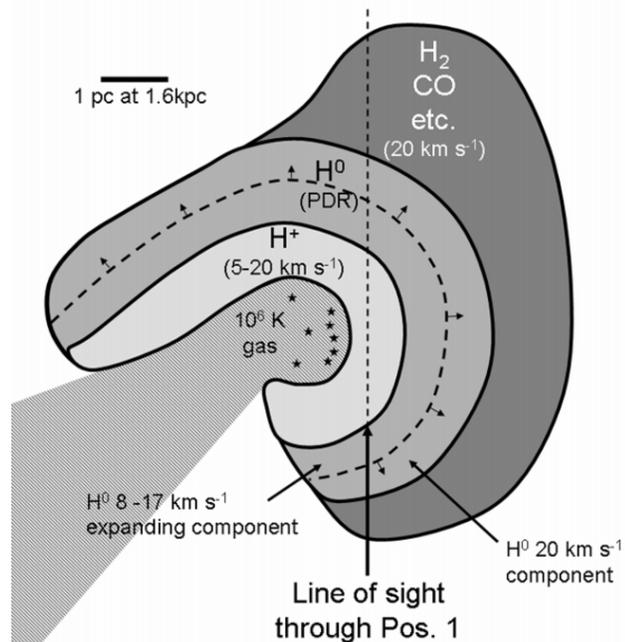

**Figure 2**. The structure that we deduce for M17, sketched as Right Ascension (horizontal axis, with W to the right) vs depth along the line of sight (vertical axis). We are looking in from the bottom, with the upward arrow showing our line of sight to Meixner92 Position 1. The known early O-type stars (Hanson et al. 1997) are marked with star symbols, at the correct horizontal locations to correspond to their Right Ascensions.

The ionizing radiation comes from a dense cluster of stars that is mostly hidden behind a tongue of obscuring material in the vicinity of the star Schulte 1. Many previous papers (cf. Meixner et al. 1992, hereafter Meixner92; Brogan et al. 1999, hereafter BT99; Brogan & Troland 2001, hereafter BT01) have interpreted M17 as an ionizing star cluster enveloped on the SW side by a bowl-shaped atomic and molecular cloud which wraps around from the SW side to cut in front of the cluster. Joncas and Roy (1986) found that the ionized gas in the NE bar is flowing back towards us at about 1.7 km s$^{-1}$ relative to the M17SW molecular cloud, and suggested that it is involved in a champagne flow from an ionized gas sheet on the far side of the star cluster.

At the position of the SW bar, we clearly see an edge-on ionization front in radio emission due to thermal bremsstrahlung in the H II region. Along the same line of sight we also see H$^0$ absorption at 21cm and OH absorption, indicating that there is significant H$^0$ and OH lying between us and the H II region. The 21 cm absorption line observations (BT99; BT01) show that the H$^0$ gas is divided into three main velocity components: a 20 km s$^{-1}$ (LSR) component which has the same velocity as the molecular cloud, an 8-17 km s$^{-1}$ blended component, and additional weak absorption spread over 0-30 km s$^{-1}$. Both the 20 km s$^{-1}$ and 8-17 km s$^{-1}$ components are measured to have much stronger magnetic fields than are typical of the interstellar medium ($B > 100\mu G$; BT99; BT01), meaning that they must be associated with the M17 complex. The underlying 0-30 km s$^{-1}$ gas has a column density which is too low for the magnetic field to be measured, but it probably is unassociated gas along the line of sight.

Figure 3, from a reanalysis of the VLA[1] data obtained by BT01, shows that the 8-17 km s$^{-1}$ component appears to be a large segment of a circular shell centered on the star cluster. The 20 km s$^{-1}$ component can be interpreted as another, concentric circular feature at a slightly larger radial distance from the stars. Such a pattern is expected for an expanding spherical shell with a systemic velocity of about 20 km s$^{-1}$ (the velocity also of the molecular gas). Since the 21 cm line is seen in absorption against the H$^+$ continuum emission, we cannot see the H$^0$ on the far side of the shell which would be expanding away from us. Figures similar to Figure 3, showing further details, can be found in BT99 and BT01.

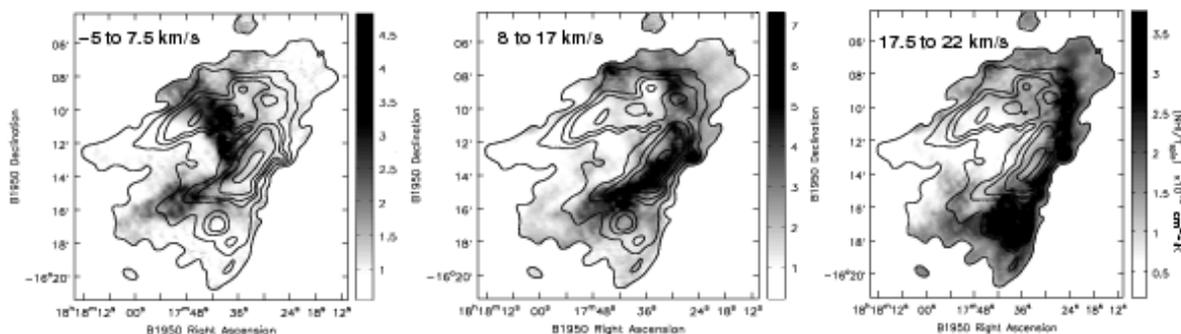

**Figure 3.** Grey-scale plot: 21 cm absorption-line maps in three velocity bins. Contour lines: 21 cm continuum emission, showing slightly different contour intervals but from the same data set as the 21 cm continuum map in Figure 1. The data are from BT01.

However, from the emission- and absorption-line velocity profiles shown in Figure 4 (measured at Meixner92 Position 1, discussed in the following subsection) it is clear that the situation is a bit more complicated. These data again are from the observations described by BT01, with the exception of the $^{13}$CO(1-0) data from Wilson et al. (1999). H110α is a recombination line seen in emission from the H II region, $\tau_{HI}$ is the optical depth profile of the 21 cm absorption line from the H$^0$ region seen against the background of the H$^+$ continuum emission, $\tau_{OH}$ is the same for the OH 1667 MHz line, and $^{13}$CO(1-0) is seen in emission from the molecular gas. Along this line of sight we see two different, distinctly separate H$^0$ and OH velocity components between us and the H$^+$ continuum source.

---

[1] The VLA is operated by the National Radio Astronomy Observatory, which is a facility of the National Science Foundation operated under cooperative agreement by Associated Universities, Inc.

The CO lines peak at a radial velocity of 20 km s$^{-1}$, so the molecular cloud and the 20 km s$^{-1}$ H$^0$ and OH components appear to be associated. The hydrogen recombination lines from the H II region include a strong component at about 7 km s$^{-1}$, while at this position the 8-17 km s$^{-1}$ component peaks at 10 km s$^{-1}$ and overlaps in velocity with the H110α profile. The differences between these velocities imply that the H$^+$ and 8-17 km s$^{-1}$ H$^0$ and OH components are expanding outwards away from the ionizing stars (moving towards us along the line of sight) relative to the H$^0$ 20 km s$^{-1}$ component and the molecular gas. We interpret this to mean that the gas in M17 is still in the process of dynamically responding to the pressure created by the newly formed star cluster, with the inner part of the neutral atomic gas snowplowing out into the outer part, as is expected on theoretical grounds (see §§6.5, 6.6 of Osterbrock & Ferland 2006, hereafter AGN3). We discuss the pressure balance in some detail in §6, and in §8 show that it is plausible that the time for the pressure disturbance to cross the PDR is greater than the age of the stars that drive the flow.

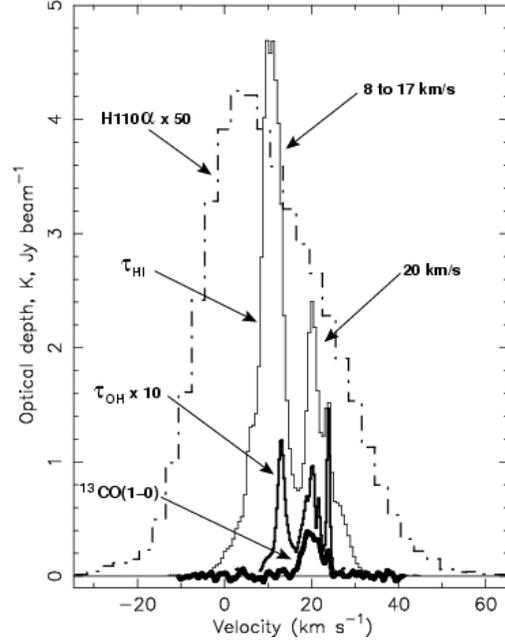

**Figure 4.** Velocity line profiles of the ionized (H110α × 50), neutral ($\tau_{HI}$),and molecular ($\tau_{OH}$ × 10 and $^{13}$CO(1-0)) gas components measured toward Position 1. Our previously unpublished H110α data comes from the VLA in its D configuration. The $\tau_{HI}$ and $\tau_{OH}$ data are both presented in BT01, while the CO data are from Wilson et al. (1999).

The radio observations measure the ratio of the neutral hydrogen column density $N(H^0)$ to the spin temperature $T_{spin}$, which is found to be $N(H^0)/T_{spin} = 9 \times 10^{19}$ cm$^{-2}$ K$^{-1}$ at the Position 1 described below. The spin temperature is in fact very poorly constrained (see the Appendix), but for assumed $T_{spin}$ in the range 150–300 K, these results indicate $N(H^0) = 1–3 \times 10^{22}$ cm$^{-2}$ for the three velocity components combined together. New in this paper are Hα flux measurements which we combine with the radio continuum flux to determine the extinction to the H$^+$ layer. We will show in §4 that for a standard gas/dust ratio, this extinction is consistent with the measured column density of neutral and molecular gas.

Figure 2 shows our understanding of the overall structure of M17. It has many similarities to Figure 6 of Joncas & Roy (1986). We represent the system as a central cluster of stars surrounded by successive layers of H$^+$, H$^0$ and H$_2$ gas to the SW side and by a background sheet of ionized and neutral gas wrapping around to the NE. The radio map shows that the edge-on part of the H II region is about 0.5 pc wide (FWHM), as projected on the sky. It is followed by the PDR containing the H$^0$, which is where the magnetic field is measured. This region is also detected in PDR tracers such as [C II] λ158μm which has a projected width of approximately 1 pc FWHM (Stutzki et al. 1988, after correcting to the distance used here). For the geometry sketched in Figure 2, these observed widths would be along a line extending horizontally to the right from the ionizing stars. The core of the molecular cloud, as projected on the sky, is 2-3 pc wide (Fig 1).

A key feature of Figure 2 is the way in which the H$^0$ region wraps around in front of the ionization front, permitting us to see the 21 cm absorption from gas in the H$^0$ region against the background continuum emission from the H$^+$ that has formed on the side towards the ionizing stars. X-ray observations (Dunne et al. 2003; Townsley et al. 2003) indicate that the region interior to the H II region is filled by hot (10$^{6-7}$ K) gas which is then flowing out to the East (the left in Figure 2).

## 2.2 A Line-of-Sight through the SW Bar

Our approach in this paper is to make a detailed model of a well studied line-of-sight through the ionization front in the SW bar, in order to understand the interplay of the recently measured magnetic fields with the various gas components in a typical PDR. We have chosen a particular position in M17 that allows us to use data from the important paper by Meixner92. That paper combined new Kuiper Airborne Observatory observations of PDR emission lines with a comprehensive review of previous data. In particular, their Table 2 draws together new and existing measurements of the surface brightness of [O I], [C II], [Si II] and CO emission lines at four different positions in the region of the SW bar[2]. The relative intensities of these lines are stated to be accurate to about 10%, but the absolute calibration of their surface brightness has an additional uncertainty of about 25%, which will affect their intensity ratio relative to Hα or the radio continuum. We will model the line of sight through Meixner92's Position 1, which is at 1950 coordinates RA = $18^h 17^m 32^s$, Dec = -16° 13' 12". Position 1 is marked with a square on Figure 1, and corresponds to the arrow on the edge-on view in Figure 2. Position 1 falls at the southern end of the arc of brightest radio emission that is seen in the high resolution 6 cm and 21 cm continuum maps made by Felli et al. (1984). In the smoothed radio map shown in Figure 1, it sits on top of the strong ridge of radio emission.

## 2.3 Magnetic Field Strength

BT99 and, at higher angular resolution, BT01 used Zeeman splitting of the 21 cm line to determine the magnetic field strength in the PDR. While the emission lines come from all along the line of sight through the nebula (modulated by the increasing extinction as we look deeper into the cloud), these magnetic field measurements and the $H^0$ column densities given in §2 are only for the absorbing column of neutral H seen against the background continuum source in the H II region. The arrow drawn on Figure 2 is solid over this absorbing region.

The line-of-sight field strength at Position 1, and for a considerable radius on the sky around it, is -50 µG. The minus sign indicates that the field is directed towards the observer. If a field of given strength is viewed from random directions, the average line-of-sight strength is one half the total field strength. Therefore, we estimate the minimum strength of the total field to be 100 µG. However, for an ordered field $B$ the component measured by the Zeeman effect is $B \cos i$, where $i$ is the angle between the field and our line of sight. For an ordered field, $B$ could be much larger than 100 µG.

The line-of-sight magnetic field peaks strongly at a value of $B_{los}$ = -600 µG at the extreme SW edge of the region which is detected in the radio continuum (see Fig. 8 of BT01). A possible reason for this (BT99; BT01) is that the field is preferentially directed along the surface of the PDR-$H_2$ boundary, so that we see it pointed towards us at the extreme limb of the PDR. Such a geometry is consistent with the near absence of far-IR linear polarization (which depends on $B_\perp$) at the same points where $B_{los}$ is highest (see BT01 Figure 19, for example). We would expect the field to be ordered in this way if the magnetic field lines have been compressed by the expansion of matter away from the star cluster, driven by both the hot shocked gas and the momentum of the starlight, as we will argue below (§7.2) is probably the case.

## 3. Stellar Content

The stellar cluster NGC 6618 illuminates the H II region of M17. Photons with energy greater than 13.6 eV are involved in the photoionization process and the excess energy is transformed into the kinetic energy of the electrons. Though there are more than 100 stars in the H II region, the ionization of the $H^+$ is dominated by only a few early O-type stars. However, the energetics of the PDR is set by the far ultraviolet continuum flux at wavelengths greater than 912Å, which is strongly influenced by late-O and B stars.

---

[2] Users of the Meixner92 paper should note that there is an error in the coordinate system shown on their Figure 1a; it is both shifted and rotated in comparison to all of the other figures in their paper. Since the other figures all agree among themselves and also with the tables of measured data given in the paper, we have assumed that they are correct.

In order to estimate the OB star content of M17, we relied on the only existing spectroscopic study of the central region (Hanson et al. 1997). The numbers are certainly incomplete, though one could hope that most of the early-O stars have been identified. We used the relation between spectral type and mass also given by Hanson et al. (1997), and binned all known stars into standard mass ranges (Massey et al. 1995). We assumed a Salpeter IMF slope of -1.35, and found the best fitting line to the data. Given that the slope was predetermined, the only degree of freedom was the intercept, which was chosen to best coincide with the two top mass bins. Here the stellar counts are most likely to be complete, though the numbers are small. Such a fit was based on stars 45 solar masses or more, and included just 5 stars. Working with the same standard mass ranges, we derived the expected stellar counts using the Salpeter fit. The estimated masses were converted back to spectral types using the relationship given by Hanson et al. (1997). The results are given in Tables 1 and 2. The use of the fitted IMF results in a non-integer number of stars in each bin (in order to correctly fit the total number of stars). These are the numbers that were used to define the incident flux in our models of the ionized gas.

Table 1
Numbers of O Stars and their Corresponding Luminosities

| Sp Range | Mass Range | # Stars | Teff | log Q (H) (s$^{-1}$) |
|---|---|---|---|---|
| O3-O4 | 70-55 | 1.25 | 47490 | 49.63 |
| O4-O5 | 55-45 | 1.40 | 45355 | 49.47 |
| O5-O6 | 45-38 | 1.56 | 43151 | 49.30 |
| O6-O7 | 38-32 | 1.95 | 41209 | 49.18 |
| O7-O8 | 32-36 | 3.03 | 39084 | 49.08 |
| O8-O9 | 26-22 | 3.15 | 36982 | 48.81 |
| O9-B0 | 22-15 | 10.26 | 34914 | 48.97 |

Table 2
Numbers of B Stars and their Corresponding Luminosities

| Sp Range | Mass Range | # Stars | Teff | Absolute Bolometric Luminosity |
|---|---|---|---|---|
| B0-B1 | 15-9.5 | 21.35 | 27700 | -8.5 |
| B1-B2 | 9.5-7.0 | 24.35 | 23700 | -7.5 |
| B2-B3 | 7.0-5.5 | 28.00 | 20350 | -6.6 |
| B3-B5 | 5.5-4.0 | 53.49 | 17050 | -6.3 |

The numbers of stars must then be converted into the total luminosity of ionizing photons. Table 1 shows values of the total Q(H) (the luminosity by number of H-ionizing photons) for each spectral range. These are based on the Q(H) per star given by Hanson et al. (1997), which are lower than many previously published values (see Vacca, Garmany & Shull 1996 for a discussion of this) but which we believe are more accurate for the reasons given in the Hanson et al. paper. The temperatures given in Tables 1 and 2 and the absolute luminosities in Table 2 are also from Hanson et al. (1997).

Adding up the contribution from stars given in Table 1 gives Q(H) = $1.35 \times 10^{50}$ s$^{-1}$. These numbers appear reasonable when one considers the values for Q(H) derived from the radio observations by Felli et al. (1984). They found Q(H) = $2.9 \times 10^{50}$ s$^{-1}$ for a distance of 2200 pc, equating to $1.5 \times 10^{50}$ s$^{-1}$ for 1600 pc.

The sum of the Q(H) values in Table 1 is greater than was estimated by Hanson et al. (1997) from the known stars in M17. This discrepancy occurs because in our analysis we are attempting to estimate the contribution from missing OB stars. The additional stars that we have inferred to be present are all at the lower end of the luminosity range: O9-B0 stars in Table 1 plus B stars listed in Table 2. These cooler stars are important because, even though the light from B stars does not appreciably add to the ionization of hydrogen, it penetrates into the PDR and greatly affects the $H_2 - H^0$ transition zone. We estimate the uncertainty in the number of late-O and B type stars to be at least 50%, because many of the B stars may still be buried in circumstellar material in such a young cluster (Hanson et al. 1997). We did not include stars later than B5 because we assume that lower mass stars will still be enshrouded in such material. Moreover, we checked that when we crudely extrapolated down to include two additional bins of lower mass, the total incident spectrum in the 500 – 10,000Å wavelength range changed only by a negligible amount, so our results are not sensitive to this assumption.

Between the uncertainty in the numbers of stars in each mass bin and in the stellar atmosphere models for each spectral type, we estimate that Q(H) is known to ±50 percent. The uncertainty in the total luminosity at λ1000Å is about a factor of two, due to the additional problems in estimating the numbers of these later-type stars.

### 4. New Observations

The existing data described above provide most of the observational constraints needed as inputs to the model which we wish to construct. But we required additional information about the gas density and extinction correction before we could proceed. The key observed parameters from both the existing and new data are listed in column 2 of Table 3. This table also contains the results from the models which we describe below in §7 and §8

Table 3
Model Results

| Parameter | Observed | Ref[a] | Model 1 | Model 2 | Model 3 | Model 4 | Model 5 | Model 2a | Model 2b |
|---|---|---|---|---|---|---|---|---|---|
| $<B>$, μG | 100-600 | 1 | 581 | 283 | 208 | 186 | 0 | 395 | 397 |
| Cosmic Ray Density/Bkg | --- | | 1 | 1 | 1 | 1 | 1 | 1 | 316 |
| $\Phi(H)$ (s$^{-1}$ cm$^{-2}$) | --- | | 1.07E13 | 1.07E13 | 1.07E13 | 1.07E13 | 1.07E13 | 1.07E13 | 1.07E13 |
| $u_{turb}$ (km s$^{-1}$) in PDR | 2-3 | 1 | 49 | 20 | 12 | 7 | 3 (fixed) | 3 (fixed) | 3 (fixed) |
| $<T_{spin}>$, K | --- | | 140 | 139 | 146 | 170 | 205 | 144 | 195 |
| Max. $N(H^0)$ in PDR, cm$^{-3}$ | --- | | 699 | 953 | 1480 | 4900 | 8910 | 935 | 943 |
| $N(H^0)/T_{spin}$, cm$^{-2}$ K | 7-10E+19 | 1 | 7.3E+19[b] | 5.6E+19[b] | 4.5E+19[b] | 3.3E+19[b] | 1.6E+19[b] | 4.3E19[b] | 5.1E+19[b] |
| $I(\lambda 6716)/I(\lambda 6731)$ | 0.98 | 2 | 0.992 | 0.977 | 0.995 | 0.985 | 0.99 | 0.98 | 0.98 |
| $S$ (Hα)[c] | 9.3 | 2 | 11.0 | 5.2 | 9.0 | 10.0 | 8.5 | 10.4 | 10.7 |
| [S II]/Hα | 0.03–0.05 | 2 | 0.056 | 0.060 | 0.064 | 0.067 | 0.068 | 0.061 | 0.061 |
| [Si II] λ34μm/Hα | 0.048 | 3 | 0.008 | 0.007 | 0.008 | 0.010 | 0.017 | 0.007 | 0.022 |
| [O I] λ63μm/Hα | 0.056 | 3 | 0.024 | 0.017 | 0.018 | 0.031 | 0.093 | 0.012 | 0.095 |
| [O I] λ145μm/Hα | 0.0075 | 3 | 0.0009 | 0.0006 | 0.0007 | 0.0015 | 0.0053 | 0.0006 | 0.0073 |
| [C II] λ158μm/Hα | 0.0099 | 3 | 0.0482 | 0.0256 | 0.0210 | 0.0176 | 0.0159 | 0.0177 | 0.0520 |
| $H^+$ thickness, pc | ≤0.5 | 1 | 1.3 | 1.2 | 0.97 | 0.85 | 0.98 | 1.2 | 1.4 |
| $H^0$ thickness, pc | ≤1 | 4 | 3.1 | 1.8 | 0.98 | 0.27 | 0.05 | 1.4 | 1.6 |
| At illuminated face: | | | | | | | | | |
| $(P/k)_{total}$, cm$^{-3}$ K | 5.0E+06 | 5 | 1.3E+08 | 5.1E+06 | 8.8E+05 | 5.6E+05 | 5.9E+05 | 5.0E+06 | 5.0E+06 |
| $(P/k)_{mag}$, cm$^{-3}$ K | --- | | 5.8E+07 | 3.6E+05 | 2.9E+03 | 1.2E+02 | 0.0E+00 | 6.4E+05 | 6.8E+05 |
| $(P/k)_{gas}$, cm$^{-3}$ K | --- | | 7.6E+07 | 4.7E+06 | 8.8E+05 | 5.6E+06 | 5.9E+05 | 4.3E+06 | 4.3E+06 |

[a] References for observations. – (1) BT99, BT01; (2) This work; (3) Meixner92; (4) Stutzki et al. (1988); (5) Townsley et al. (2003).
[b] After multiplying by $2^{1/2}$ to account for the non-radial direction of the observed absorption path.
[c] In units $10^{-13}$ erg cm$^{-2}$ s$^{-1}$ arcsec$^{-2}$, dereddened.

### 4.1 Optical long-slit spectroscopy

Long-slit spectra of M17 were taken on 31 October 1990 at CTIO, using the RC spectrograph at the 1.5m telescope. The spectral range was 6190-7463Å. The slit width was 4 arcsec with a length of 450 arcsec.

The resolution, as measured from a star in the slit, is 3.2 arcsec FWHM, sampled at 1.9 arcsec per pixel. The two slit positions used here were both with the slit in PA 58°. They overlapped in the region of the O5 V star, Schulte 1 (listed as star B111 in Hanson et al. 1997), which was placed in the slit and used as a positional reference (Fig. 1). These slit positions are perpendicular to the ionization front and the PDR. Basic data reductions, through wavelength and flux calibration, were made using IRAF. The spectra were extracted into 10 arcsec bins chosen so that one of these bins was centered on Schulte 1. The flux calibration was obtained from observations through a wide slit of three different standard stars from the Baldwin & Stone (1984) list. The M17 spectra were not sky-subtracted because we are interested here only in the strengths of the Hα and [S II] λλ6717, 6731 emission lines, which are not contaminated by night-sky lines.

**4.2 Density Measurements**

We used the [S II] λλ6717, 6731 doublet ratio from these optical spectra to measure the density in the H II region as a function of position along the slit. The strengths of the [S II] lines were measured automatically using a program (graciously supplied by A. LaCluyzé) which fit the continuum with a first order polynomial and the emission lines with Gaussians. The output was inspected for quality and remeasured manually with the IRAF routine SPLOT on a case-by-case basis when needed. The results are shown in Figure 5a. The spectrograph slit passed within 9 arcsec of Meixner92's Position 1, well within the beam width (~30 arcsec FWHM, or greater) of the various far-infrared and mm-wave observations that were compiled by Meixner92. However, due to a patch of obscuration (which is easily seen on 2MASS infrared images, as in Figure 1 of Wilson, Hanson & Muders (2003) there is a large dip in the [S II] surface brightness and the lines became too faint to measure over the 30 arcsec where the slit is closest to Position 1. Since the [S II] line ratio is nearly constant on either side of this region, we used the average ratio over 30 arcsec bins to either side of the obscured range (see Fig. 5a). The observed ratio is I(λ6717/λ6731) = 0.98±0.05, corresponding to an electron density $n_e$ = 560 cm$^{-3}$ (AGN3).

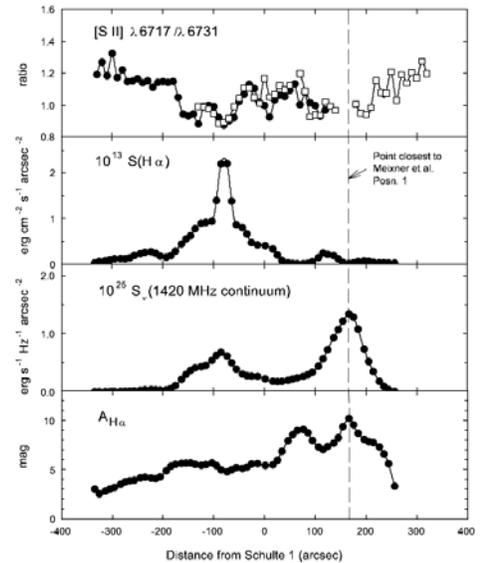

**Figure 5.** Results from long-slit spectroscopy and comparison to 21cm data of BT01. From top to bottom: (a) The [S II] doublet ratio (the two different symbols indicate our two overlapping slit positions); (b) the Hα surface brightness along the slit after spatial smoothing to match the radio resolution; (c) the 21 cm surface brightness extracted along the same line as the optical slit position; and (d) the derived extinction at Hα.

Table 3 also lists [S II]/Hα, the intensity ratio of the sum of the [S II] λλ6717, 6731 lines to Hα. The observed value at Position 1 was estimated by measuring this ratio in the regions to either side of the obscured patch. There is a gradient, with the larger ratio on the side farther away from the ionizing stars. The range is indicated in Table 3.

**4.3 Extinction**

The long-slit optical observations also provide information about extinction due to dust, because we can compare the Hα surface brightness measured from them to the 21cm thermal bremsstrahlung emission which comes from the same ion in the same gas. While Hα emission is obscured by dust, the long wavelength radio continuum is essentially unaffected. The 21cm data (BT01) are from the VLA and have spatial FWHM resolution of ~26 arcsec. The SW bar therefore is fully resolved along its length and marginally resolved in the perpendicular direction. Fortunately, Hα is still bright enough to be measured at Position 1. We smoothed the optical Hα to the same spatial resolution as the radio data (Figures 5b-c) and found the Hα/radio intensity ratio at each point along the slit.

For a free-free continuum due to 90 percent $H^+$ and 10 percent $He^+$, the intrinsic ratio of $F_{H\alpha} / \upsilon F_\upsilon$ (1420Mhz) is $4.37 \times 10^4$ (AGN3). The scattering part of the extinction behaves differently for an extended source such as M17 than for the stellar case (AGN3 §7.6). After a correction for that effect, the observed ratio implies that at H$\alpha$ the maximum extinction along our slit is 9.46 mag. Hanson et al. (1997) have determined the reddening to a number of different stars in M17 and find similar maximum values, in the range 10-15 mag. Our maximum occurs at the point along the slit closest to Position 1. This extinction value is directly measured, but its conversion to $A_V$ does depend slightly on what reddening law is used. The spectroscopy presented by Hanson et al. (1997) suggests that $R$ varies with $A_V$ in the sense that $R$ is larger in more heavily shielded regions. Such a variation is consistent with the idea that grains grow by a variety of chemical processes in cold shielded regions (Massa & Savage 1984; Cardelli & Clayton 1991). This paper concerns the conditions within the PDR with its large hydrogen column density. Accordingly we use the large value of $R$ observed in the most heavily obscured regions of M17 and also found in Orion and the Carinae star forming region. Specifically, we use the reddening curve for the Orion Nebula given in Table 7.1 of AGN3, which has $R = 5.5$.

The resulting visual extinction (Fig. 5d) varies from $A_V \sim 1-2$ at the NE end of the slit, through $A_V \sim 4$ in the NE bar, and reaches $A_V \sim 11$ in the SW bar next to Position 1. In this same line of sight, BT01 measured $N(H^0)/T_{spin} = 9.0 \times 10^{19}$ cm$^{-2}$, corresponding to $N(H^0) = 1.3 \times 10^{22}$ cm$^{-2}$ for $T_{spin} \sim 150$K (see the Appendix). For the standard relation $N(H) = N(H^0) + 2 \times N(H_2) = 2 \times 10^{21} A_V$ (Bertoldi & McKee 1992), we expect $A_V \sim 7$ mag total extinction towards the SW bar due just to the dust that is expected to be mixed in with the observed $H^0$. Allowing for 3 mag further extinction in overlying molecular gas, the extinction predicted from the observed $N(H^0)$ is consistent with the $A_V$ derived from the emission lines, indicating that the [S II] lines at the position of the SW bar probably do in fact measure the gas density *at the ionization front*.

The extinction properties discussed above are for the total obscuration between us and the H II region. The 21 cm column density profile towards Position 1, shown in Figure 4, includes diffuse wings which are likely caused by foreground interstellar absorption. We estimate that about 2/11 of the column density is in a very broad component underlying the sharper absorption due to the M17 SW gas, and that this contributes $A_V \sim 2$ mag to the total extinction. A foreground contribution of this size is reasonable given that the stars in the M17 cluster with the lowest extinction have $A_V = 3$ mag. Subtracting 2 mag of foreground extinction leaves an $A_V = 9$ mag of internal extinction within M17, along the line of sight to Position 1.

## 5. A Model of M17

We use version C06.04.20 of "Cloudy", the equilibrium photoionization code (Ferland et al. 1998), to model the line of sight through Position 1. The following subsections describe the specific ways in which the more important observational constraints were applied to the Cloudy models, except that the key issue of pressure support is given its own section (§6).

### 5.1 Geometry

The geometry derived from the observations of M17 was discussed in §2. Although the actual nebula is at best an incomplete shell, we approximated it in the calculation with a full spherical geometry. Therefore we have the exciting stars surrounded by the layers of the H II region, the PDR region and the molecular cloud. The calculation of various physical parameters was carried out starting from the illuminated face and moving radially outwards, away from the ionizing photon source and towards the molecular cloud. The Cloudy simulation corresponds to a ray perpendicular to the line of sight, extending from the ionizing star into the cloud. As is sketched in Figure 2, the observed line of sight to Position 1 cuts through this spherical geometry at an angle, which increases the total path length and hence the total line strengths and the total extinction.

## 5.2 Incident Radiation Field

The shape and intensity of the incident continuum are provided as input parameters for the model. We use the COSTAR (Schaerer & de Koter 1997) stellar continua for spectral classes O3 - O9 and ATLAS (Kurucz 1991) models for later spectral classes.

Although we infer that there are many late O and early B stars at unknown positions within M17, most of the ionizing continuum radiation must come from the 8 known stars of type O5 and earlier that lie, at least in projection, within the central cavity. Figure 2 of BT01 shows the positions of these stars relative to the H II region. Most of these stars project to a position rather close to the SW bar, as is sketched here in Figure 2. Giving equal weighting to the light from each of these stars, we find that the centroid of their light comes from a point that projects to only 1 arcmin, or 0.46 pc, from the peak of the radio continuum emission in the SW bar. Allowing for the 0.5 pc FWHM thickness of the $H^+$ zone, this would correspond to a light source lying only 0.23 pc from the illuminated face of the H II region. However, since we don't know where these stars lie along the line of sight, we assumed that the true incident flux is two times smaller than the value that would come from the closest possible projected distance. We also assumed that the incident flux from the (unseen) B stars should scale in the same way. The above radiation field corresponds to an incident flux of ionizing photons $\Phi(H) = Q(H)/4\pi r^2 = 1.07 \times 10^{13}$ $s^{-1}$ $cm^{-2}$.

X-rays contribute to the heating and ionization of the gas. The ROSAT results of Dunne et al. (2003) indicate a total diffuse X-ray luminosity of $2.5 \times 10^{33}$ ergs $s^{-1}$ in the ROSAT PSPC 0.1-2.4 KeV band at a mean characteristic temperature of $kT \sim 0.72 KeV$ or $T \sim 8.5 \times 10^6$ K. The ionization and chemical effects of the X-ray continuum are fully included in our model, as described by Abel et al. (2005). For purposes of calculating the flux incident on the illuminated face of the cloud, we placed the x-ray source at the same position as the centroid of the starlight.

Our calculations also include the energy input from ionization and heating by cosmic rays. The cosmic-ray density at the location of M17 is unknown, but the Galactic background density is of the order of $2.6 \times 10^{-9}$ $cm^{-3}$ (Williams et al. 1998), which is the value we initially assumed for our models. However, we investigate the effect of higher values in §8.

## 5.3 Chemical Composition

Although our Cloudy models include the $H^+$ and $H_2$ zones as well, our main focus is on the structure of the PDR. Tsamis et al. (2003) and Esteban et al. (1999) have measured the elemental abundances from emission lines from the $H^+$ zone. For He, N, O, Ne, S, Cl, Ar and Fe, we used the results from Table 11 of Esteban et al. (1999) for $t^2 = 0$ at their Position 3 (which is in the relatively unobscured NE bar, roughly 4 arcmin NE of Meixner92 Position 1).

However, Esteban et al. find that the C/O abundance ratio in the $H^+$ zone is C/O = 1.7. The large C/O ratio found for the $H^+$ zone cannot apply for the molecular gas. If the C/O ratio was that high in the molecular cloud, where CO has fully formed, the less abundant O would be fully depleted, leaving free C. The molecular cloud would have a C-dominated chemistry rather than the observed oxygen rich gas, resulting in C-bearing minor molecules, not O-bearing as observed. We know of no independent measurements of C/H in the molecular cloud. There is also no evidence that it is unusual. Our assumption is that the C/O ratio, and probably also the C/H abundance, is considerably lower in the $H_2$ zone than in the $H^+$ zone. Bergin et al. (1997) find that an initial abundance of $n(C^+)/n(H_2) = 1.46 \times 10^{-4}$ produces a chemistry similar to that seen in M17. That $n(C^+)/n(H_2)$ ratio corresponds to a C/H ratio of $7.3 \times 10^{-5}$, which we find below reproduces the observed [C II] 158 μm line luminosity to within some reasonable level of agreement. We use this abundance in our models.

For the remaining chemical elements, we use the standard values included in Cloudy for H II regions, most of which are derived from the study of the Orion Nebula by Baldwin et al. (1991). Among these is Si/H = $4.07 \times 10^{-6}$, which affects the comparison we make below to the observed [Si II] 35 μm line strength. We will discuss these abundances further in §8.5.

We must also include the presence of dust grains since they play an important role in establishing the overall equilibrium of the gas in photoionized clouds. They not only provide extra heating but also lead to the gas phase depletion of the refractory elements. These effects will cause the gas to equilibrate at a higher temperature and absorb ionizing radiation. We include grains similar to the ones in the Orion Nebula (Baldwin et al. 1991). These are large-R grains with a fairly gray ultraviolet extinction. Abundant and ubiquitous in their presence, the polycyclic aromatic hydrocarbons (PAHs) are known to be efficient in heating the gas and aiding the formation of hydrogen. We use the Bakes & Tielens (1994) PAH size distribution with ten times the default abundance (appropriate for Orion), and with a total PAH abundance that is proportional to $N(H^0)/N(H_{total})$. The gas to dust ratio is 180, which is just above the ISM value of 140.

### 5.4 Density

Another input value for the models is the gas density at the illuminated face of the cloud. The actual measured value is the [S II] intensity ratio $I(\lambda 6717)/I(\lambda 6731)$, which corresponds to an electron density somewhat lower than that at the illuminated face, with some weighting by any variation in $N_e$ through the $S^+$ ionization zone. Since Cloudy computes the intensities of each of the [S II] lines, the input density was iterated until the observed line ratio was matched.

### 5.5 The Cloud's Outer Boundary

We used an extinction value of $A_V = 6.36$ as the stopping criterion to simulate the line of sight to Position 1. This $A_V$ is the observed extinction internal to M17, $A_V = 9$ mag, multiplied by $\cos(45°)$. It accounts for the fact that while Cloudy models a radial ray outwards from the ionizing stars, we observe through a longer line of sight (see Fig. 2).

## 6. Pressure Support in a Gas Cloud Reacting to Stellar Energy Input

A key issue in our models is the physical source of the pressure in the gas, and the way in which the pressures due to various sources play off against each other. This section explains in some detail the approach that we have taken, which is significantly different from assuming either constant pressure or constant density throughout the cloud.

### 6.1 A Time-Steady Hydrostatic Cloud

Interstellar gas near a newly formed star cluster is a dynamical, evolving environment (Henney 2006; Henney et al. 2005). As the cluster forms and stellar winds and radiation pushes back into surrounding gas, the gas tends to approach rough hydrostatic equilibrium. By "hydrostatic equilibrium" we mean that within each volume element the forces caused by various pressures within the system balance. Although there is pressure balance at every point within the gas, hydrostatic equilibrium does not necessarily require that the total pressure is constant throughout the cloud, as we will discuss in detail in §7.2. Although an H II region with associated PDR never reaches a true hydrostatic condition, this simplification does give a good description of the innermost regions of the Orion H II region (Baldwin et al. 1991; Wen & O'Dell 1995). We pursue such a model here, as a reasonable approximation of the true dynamical situation in M17.

The star cluster with its radiation field and winds is the energy source for the environment. In a time-steady model, the force caused by the outward momentum in starlight is balanced by pressures present in the nebula, including gas, magnetic, and turbulent pressure. This balance is close to what is observed.

### 6.2 The Star Cluster & X-Ray Bubble

The direct effect of the stellar wind is the pressure support it provides for the illuminated face of the $H^+$ region. Stellar winds create hot gas by forming a shock as the wind strikes the nebula. The high-speed wind will be thermalized by the shock and will not penetrate past the hot gas/nebular gas interface. X-ray observations of the M17 region show a bubble of hot gas surrounding the star cluster and interior to the H II region. The pressure of the x-ray emitting gas is $P_x/k = 5\times10^6$ cm$^{-3}$ K (Townsley et al. 2003).

**6.3 The Nebula**

The nebular gas in the H II region is in pressure balance with the hot gas. The gas pressure in the H II region can be estimated using the optical observations discussed above. The electron density measured with the [S II] lines, $n_e \sim 560$ cm$^{-3}$, combined with a typical H II region temperature $T_e \sim 9000$ K, corresponds to $P_{neb}/k = 5 \times 10^6$ cm$^{-3}$ K, equal to $P_x$. Thus the hot gas provides the support that holds the H II region in place.

**6.4 The PDR and Its Magnetic Field**

As was noted above, extensive maps of the line of sight magnetic field in the PDR adjoining the H II region exist (BT99; BT01) with a range 50 – 600 μG (§3.2). The pressure due to the magnetic field is $P_B/k \sim B^2/8\pi k$ so the magnetic pressure corresponding to B = 100 μG is

$$P_B / k \approx \frac{B^2}{8\pi k} \approx (B_{100\mu G})^2 2.9 \times 10^6 \text{ [cm}^{-3}\text{ K]}, \qquad (1)$$

similar to the gas pressures mentioned above. The general ISM is observed to have turbulent and magnetic energy densities that are on average in equipartition. Numerical simulations suggest that such equipartition is true in general although the relationship is not fundamental, and may not apply to any particular parcel of gas (Mac Low & Klessen 2004). The fact that the line widths indicate supersonic motion suggests that the turbulence is due to some sort of MHD wave (Heiles & Troland 2005). For the equipartition case the total of turbulent and magnetic pressures would be twice the value for just $P_B$.

Magnetic field effects are included in the model by specifying the field strength and geometry at the illuminated face. We then must assume a scaling law between magnetic field strength and gas density in the sense $B \propto n^{\gamma/2}$ (Henney et al. 2005). This same law is often expressed as $B \propto n^\kappa$ in the literature dealing with magnetic fields (see §8.2 of Heiles & Crutcher 2005). Such a law assumes that the magnetic field is coupled to the gas (flux freezing). The value of $\gamma$ depends upon the geometry of the field and the nature of the process bringing about the increase in density. If a slab of gas is compressed in one dimension with the field parallel to the slab, as for example in a shock front where the field lies parallel to the front, $\gamma = 2$. If a spherical cloud contracts in three-dimensions, $\gamma = 4/3$. The latter case corresponds to a gravitationally contracting cloud within which magnetic field energies are much smaller than gravitational energies. If the gas moves along the field in one dimension, then $\gamma = 0$ (e.g. shock front with magnetic field perpendicular to the shock front).

For large samples of molecular clouds, Basu (2000) found that $B \propto \sigma_v n^{0.5}$, where $\sigma_v$ is the velocity dispersion within the cloud. This is the expected relation for clouds which are flattened along the direction of the magnetic field by self-gravity.

For M17, we ran models with three different values: $\gamma = 2, 4/3$ and 1. The results are fairly similar. We show only the $\gamma = 2$ models below, which correspond to a situation that would produce strong magnetic fields directed parallel to the surface of the PDR. A magnetic field configuration of this type would occur in a situation where winds and light from the star cluster push back surrounding gas and as a result amplify the field. Under those circumstances the highest projected field strength would be seen at the very edge of the PDR, as is observed in M17 (§2.3). We ran models that produced a series of increasing average field strengths, in order to explore the effects of this key parameter. The input parameter which we varied was the field at the illuminated face of the H II region.

## 6.5 The Final Pressure Law

We assume hydrostatic equilibrium in relating the x-ray, nebular, PDR, and molecular regions. At any particular position in the cloud, the sum of the contributions from the gas pressure, magnetic pressure, turbulent pressure, line radiation pressure and the outward pressure of starlight, remains a constant;

$$P_{tot}(r) = P_{tot}(r_o) + \int a_{rad}\, r\, dr \qquad (2)$$
$$= P_{gas} + P_{lines} + P_{continuum} + P_{ram} + P_{turb} + P_{magnetic}$$

Here $a_{rad}$ is the photon momentum absorbed per unit time per unit mass at each point in the gas, and the integral is from the illuminated face out to the point $r$. Line widths in the range 3–6 km s$^{-1}$ are seen in the 21 cm line profile (see Fig. 4, and also BT99). Turbulence serves as a line broadening mechanism and contributes a component of turbulent pressure. In most of the calculations presented below we assume that magnetic and turbulent pressures are in equipartition and so simply double the magnetic pressure. Our models assume that the cloud is static, so $P_{ram} = 0$.

## 7. Results: Magnetically Supported PDRs

### 7.1 Dependence on Magnetic Field Strength

Although the ability to measure the line-of-sight component of the magnetic field in the PDR makes M17 an almost unique object for the sort of study presented here, the conversion of the measured line-of-sight $B$ component back to the average field $<B>$ depends heavily on the actual geometry and ordering of the field (§2.3). Depending on the geometry, values of $<B>$ in the range 100-600 µG are all consistent with the observed $B_{los}$ = -50 µG at Position 1. We varied the strength of the magnetic field at the illuminated face, so that $<B>$ would span the range from large but still plausible values, down to $<B>$ = 0. The results are shown as Models 1-5 in Table 3 and in Figure 6.

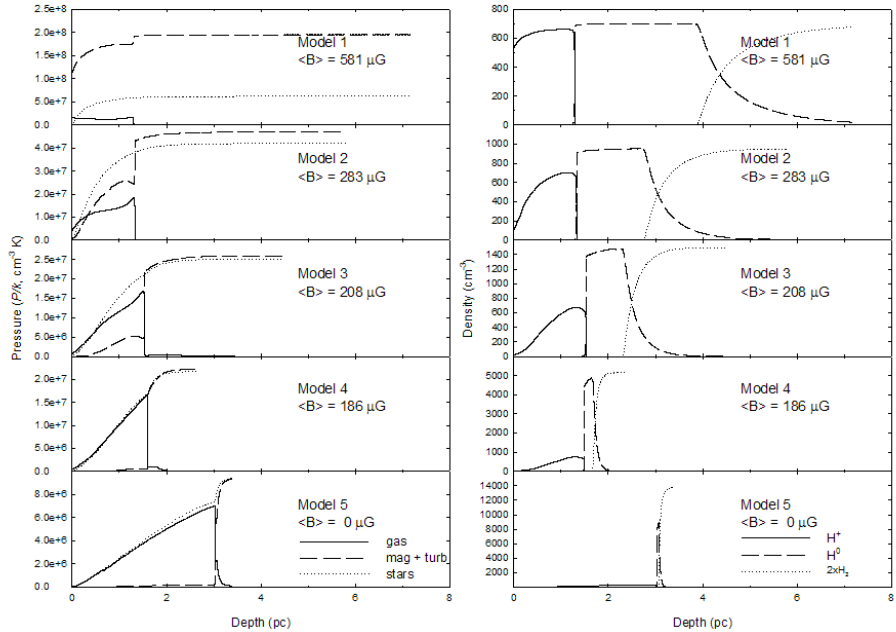

**Figure 6.** Pressure and density as a function of depth, for a series of models with decreasing mean magnetic field $<B>$. These models all have $\gamma = 2$, $F = 3$ (see §6.3), the default incident flux, equipartion between the magnetic and turbulent pressures, and are stopped when $A_V$ = 6.36 is reached. The plots are on linear scales to emphasize the great difference in the appearance on the sky of gas with a strong magnetic field from that with no magnetic field. The dotted line in the left-hand plots is the total integrated pressure at each depth due to the momentum absorbed from starlight, discussed in §6.2.

The geometry and properties of M17 are nearly fully constrained by observations. The major remaining uncertainties are the density and magnetic field at the illuminated face of the cloud. These are presented in Table 3 in terms of pressure at the illuminated face; the magnetic pressure is just $B^2/(8\pi)$ and the gas pressure is proportional to $n_H$ since the temperature is nearly the same from model to model. We assumed energy equipartition between the turbulent and magnetic pressure in this series of models, so there is also a turbulent pressure at the illuminated face that is equal to the magnetic pressure, with $P_{total} = 2P_{mag} + P_{gas}$. Whether or not equipartition exists has little effect on our basic conclusions, but we will reconsider it below. The observed quantity listed for $(P/k)_{total}$ at the illuminated face is the pressure of the x-ray emitting gas (Townsley et al. 2003). $I(\lambda 6716)/I(\lambda 6731)$ is the [S II] intensity ratio. So that models with different physical parameters could be compared on equal footing, the density of each model was iterated to reproduce the observed [S II] ratio to within 1.5% percent. $<B>$ is the average magnetic field, weighted by $N(H^0)/T_{spin}$, which is proportional to the $H^0$ opacity at each point along the line of sight. The average spin temperature, $<T_{spin}>$, is calculated from $<1/T_{spin}> = \int [n(H^0)/T_{spin}]\, dr\, /\, N(H^0)$. The thicknesses of the $H^+$ and $H^0$ regions are measured between the points where the ion density is half of its maximum value.

Table 3 also lists the surface brightness $S(H\alpha)$. The observed value has been dereddened. The values given for the models are for a total observed path length through the $H^+$ zone that is 3 times larger than the radial path calculated in the models, as is suggested by the sketch in Figure 2. Also listed are the observed and calculated strengths of four strong PDR emission lines: [Si II] $\lambda 34\mu m$, [O I] $\lambda 63\mu m$, [O I] $\lambda 145\mu m$ and [C II] $\lambda 158\mu m$. The observed values are from Meixner92.

Turning to Figure 6, the left-hand column shows the various pressures as a function of depth into the cloud. $P_{stars}$ is the integrated radiation pressure due to the absorption of the momentum in the cluster's starlight, and will be discussed in the following section along with the reasons for the behaviors of these different pressures as a function of depth. The right hand column shows the densities of hydrogen atoms in the $H^+$, $H^0$ and $H_2$ forms, again as a function of the depth into the cloud.

It is clear from Models 1-5 that the presence of any appreciable magnetic field drastically increases the physical thickness of the $H^0$ zone (the PDR). The thickness is set by the path length needed to reach a given column density such that the photons capable of disassociating the molecular gas are absorbed. When there is a large amount of magnetic pressure to support the gas, the gas pressure needed to maintain pressure balance is much lower. The result is that as the initial magnetic field is increased the gas density is much lower than it would be without the field, resulting in a much longer photon path length through the PDR.

## 7.2 Relationship between $Q(H^0)$ and B.

Figure 6 shows that the physical size of the PDR increases as the amount of magnetic pressure increases. The panels on the left identify various contributors to the total pressure. In the weak field case the cloud is supported entirely by gas pressure, the densities are high, and the cloud is much smaller than observed. As the magnetic field increases the magnetic pressure does too - most of support is given by the field in most geometries presented.

Our models show that the pressure in the PDR and molecular gas is dominated by magnetic pressure. The average magnetic field in these regions rises to similar high values even though the field strength at the illuminated face of the H II region is given various much smaller values. The field strength always equilibrates at about the same value because the magnetic pressure in the PDR is balancing the net outward force produced by the star cluster. This situation is most obvious in the 1-3 o'clock positions in Figure 2, where the PDR is sandwiched between the (immovable) molecular cloud and the star cluster. The outward force produced by the star cluster is the sum of the gas pressure in the hot gas detected by Townsley et al. (2003) and the force produced by absorption of the stellar radiation field.

Here we find a relationship between $Q(\mathrm{H}^0)$, the number of ionizing photons emitted by the star cluster per second, and the magnetic field $B$, assuming hydrostatic euilibrium. We assume that most of the momentum in the stellar radiation field is in ionizing photons. The total momentum per unit time is then $Q(\mathrm{H}^0)\langle h\nu \rangle / c$, where $\langle h\nu \rangle$ is the mean photon energy of an O star and $c$ is the speed of light. If the hydrogen ionization front occurs at $R_\mathrm{H}$ away from the star cluster then the momentum per unit area and time, the pressure, pressing on the $\mathrm{H}^0 - \mathrm{H}^+$ ionization front is

$$P_{stars} = \frac{Q(\mathrm{H}^0)\langle h\nu \rangle}{4\pi R_\mathrm{H}^2 c} \quad . \tag{3}$$

$P_{stars}$ is plotted on Figure 6, to show how it builds up with depth into the cloud. The total pressure pushing on the gas is $P_{stars}$ plus the pressure due to the x-ray emitting hot bubble, $P_{tot} = P_x + P_{stars}$.

The total pressure must be balanced by the magnetic field, or

$$P_x + P_{stars} = P_x + \frac{Q(\mathrm{H}^0)\langle h\nu \rangle}{4\pi R_\mathrm{H}^2 c} = \frac{B^2}{8\pi} \quad . \tag{4}$$

The balancing magnetic field is then

$$B^2 = 8\pi (P_x + P_{stars}) = 8\pi P_x + \frac{2Q(\mathrm{H}^0)\langle h\nu \rangle}{R_\mathrm{H}^2 c} \quad . \tag{5}$$

Inserting typical values, $nT_x \sim 10^6$ cm$^{-3}$ K, $\langle h\nu \rangle \sim 15$ eV, $Q(\mathrm{H}^0) \sim 10^{50}$ s$^{-1}$, $R_\mathrm{H} \sim 1$pc, we find

$$B^2 = 3.5\times 10^{-9} nT_{x6} + 1.7\times 10^{-8} \frac{Q_{50}\langle h\nu \rangle_{15}}{R_1^2} \tag{6}$$

where $T_{x6}$ is in units of $10^6$ K, $Q_{50}$ is in units of $10^{50}$ s$^{-1}$, $<h\nu>_{15}$ is in units of 15 eV, and $R_1$ in units of 1 pc. Converting to μG,

$$B_{\mu G} = 59\sqrt{nT_{x6} + 4.8\frac{Q_{50}\langle h\nu \rangle_{15}}{R_1^2}} \ \mu\mathrm{G} \tag{7}$$

For $R_1 = 0.32$, $nT_{x6} = 3.5$, $Q_{50} = 1.35$ and $<h\nu>_{15} = 1$ as we have used here for M17, $B = 487$ μG, close to the peak $B$ measured by BT01. The agreement between the calculated and observed values shows that the magnetic field has been compressed until its pressure balances the pressure from the hot x-ray emitting gas and the absorbed starlight.

The above assumes that the turbulent pressure is negligible, as is the case in our best model described below. For the equipartition models shown in Figure 6, where $P_{turb} = P_{mag}$, the value of $B$ found from Equation (7) would include the turbulent pressure, and so would be $2^{1/2}$ times larger than that calculated by Cloudy.

The pressures calculated in our Cloudy models (left-hand panels of Figure 6) closely follow the behavior described above. $P_{stars}$ starts out as zero, and then builds up throughout the H$^+$ zone and into the PDR. The magnetic pressure increases in the PDR and molecular gas to balance the integrated radiation pressure. However, as is conspicuous in models 1 and 2, $P_{stars}$ is always less than the sum of the other pressures. The difference is due to the pressure at the illuminated face (the sum of $P_{gas} + P_{mag} + P_{turb}$ at the point where the depth is zero). If the residual pressure is not balanced by an external pressure, the ionized gas will expand off the illuminated face.

### 7.3 Dependence on Other Parameters

The models presented in Table 3 and Figure 6, including Models 2a and 2b which we describe below, all used $\gamma = 2$ in the $B \propto n^{\gamma/2}$ relation. This value of $\gamma$ is appropriate for the geometry in which the magnetic field is parallel to the surface of the PDR and is being compressed perpendicular to the field lines by the momentum of the starlight, which is also the geometry consistent with the high $<B>$ in our best model. However, we also ran parallel grids of models using $\gamma = 4/3$ and $\gamma = 1$. These gave results similar to those of the $\gamma = 2$ models. Comparing models for the three values of $\gamma$ which produced the same $(P/k)_{total}$ at the illuminated face, the intensity ratios of PDR lines relative to Hβ (discussed below in §8) changed by 17 percent at most. The average magnetic field varied by between 7 and 13 percent from the $\gamma = 2$ models. Although $(P/k)_{total}$ at the illuminated face was the same in each case, its distribution between its different components did change. The gas pressure was about 25 percent higher for the $\gamma = 4/3$ and $\gamma = 1$ cases than for $\gamma = 2$. The initial magnetic pressure varied by a much larger amount but is at most 20% of the total pressure at the illuminated face. Since the different models all produce similar average magnetic fields in the PDR, such a variation is not important. Finally the thickness of the PDR and HII regions get smaller as $\gamma$ decreases since the density must increase more to balance the pressure. For $\gamma = 4/3$ and $\gamma = 1$, the thicknesses change by about 18 and 50 percent respectively.

Another parameter that required an arbitrary choice is the scaling factor between the turbulent pressure and the mean kinetic energy due to turbulence. We can write (Heiles & Troland 2005)

$$P_{turb} = \frac{F}{6} n u_{turb}^2 \qquad (8)$$

where $n$ is the gas density and $u_{turb}$ is the turbulent velocity. For the models in Figure 6 we used $F = 3$, which is appropriate for isotropic turbulent motions. However, $F = 2$ is the proper value for turbulent velocities that are perpendicular to the magnetic field, such as Alfven waves. Since the models in Figure 6 assumed that $P_{turb} = P_{mag}$, the only effect of changing $F$ would be to change the calculated values of the turbulent velocity. However, if we had instead been trying to match an observed turbulent velocity, then the calculated $P_{turb}$ would have decreased, and $P_{mag}$ would have increased to take up the slack.

## 8. A "Best" Model at Position 1

### 8.1 Overall pressure balance with the hot x-ray emitting gas

We now fit a magnetically supported model to the observed parameters for Position 1 in M17. We rely also on measurements perpendicular to our line of sight, so we implicitly assume that there is at least some rough spherical symmetry. The final model will have many adjustable parameters, and may not be unique. However, it demonstrates that a simple hydrostatic geometry can reproduce the large scale properties of the M17 complex. The remaining parameters are the turbulence, cosmic rays (which determine the chemistry), and the individual abundances in the PDR (which may be different than in the $H^+$ zone because of depletion onto dust).

As was described in §6, the observations directly show that the gas pressure in the hot x-ray component and in the H II region are both approximately the same as the magnetic pressure in the PDR. We have therefore computed models in which the $H^+$, $H^0$ and molecular gas are all in pressure equilibrium (once we have included the momentum absorbed from the starlight). The remaining piece to have the whole M17 structure in pressure equilibrium is for the total pressure (the sum of the magnetic, turbulent and gas pressures) in the H II region at its illuminated face to be equal to the pressure of the hot x-ray gas that is in contact with that illuminated face. Model 2 (with $<B> = 283$ μG) achieves this pressure balance.

### 8.2 Turbulent velocity

Model 2 is in pressure balance with the x-ray emitting gas, but we assumed that the magnetic and turbulent pressures are in equipartition. We can test this assumption by calculating an expected velocity broadening of the emission and absorption lines. Model 2 produces $u_{turb}$ =20 km s$^{-1}$. The H$^0$ and CO line profiles shown in Figure 4 are fully resolved, and have FWHM in the range 3-5 km s$^{-1}$, corresponding to a 3-dimensional velocity dispersion $u_{turb}$ = 2-3 km s$^{-1}$. The fact that the observed $u_{turb}$ is much smaller than the computed value suggests that the magnetic and turbulent pressures may not yet have come into equipartition. Non-equilibrium is known to occur elsewhere in the ISM, for instance in Orion's Veil (Abel et al. 2006).

Therefore, as an opposite extreme, we calculated Model 2a with the turbulent line widths fixed at the observed value, $u_{turb}$ = 3 km s$^{-1}$. The results are listed in Table 3. As expected, they are extremely similar to those for Model 2 except that <B> increased by a factor 1.4 (to 395 μG), because the turbulent pressure became negligible so that $P_{mag} = B^2/(8\pi)$ approximately doubled in order to maintain pressure balance. Using the observed $u_{turb}$ brought the computed <B> closer to the observed peak <B> value. The H$^0$ region also became slightly narrower.

### 8.3 PDR Line Strengths

Model 2a still fails to reproduce two of the observations. The first is the strength of the PDR emission lines relative to Hα or to other lines from the H II region. Table 3 lists the observed and predicted strengths of the most important PDR cooling lines ([C II] 158μm, [O I] 63, 145μm). These lines are up to a factor of 10 different in the model in comparison to the observations by Meixner92.

The second problem area concerns the observed OH absorption line shown in Figure 4, which shows a velocity component of OH mixed with the 8-17 km s$^{-1}$ expanding shell (Fig. 4), implying that OH and H$^0$ are at least partially mixed together. In the above model, the H$^0$ and OH distributions do not overlap.

### 8.4 Cosmic ray flux

The problems with the PDR line strengths may indicate that much of the PDR emission comes from high density clumps, as proposed by Meixner92, which are not included in our model. We return to this point below in §9. Here we first explore an alternate possibility which involves cosmic ray heating. For lack of alternate information, the models described so far have used the Galactic background cosmic ray flux. However, it is entirely possible that the compression of the magnetic field in the PDR has carried along the cosmic rays, leading to a much higher cosmic ray density. In fact, equipartition between the cosmic ray and magnetic field energy densities is observed in the local ISM (Webber 1998), and is thought also to occur on the scale of radio lobes in radio galaxies (cf. Burbidge 1959). If equipartition were to occur, the cosmic ray density in the region where <B> = 395 μG would be 2050 times greater than the background value we have assumed so far.

We varied the cosmic ray flux in our model to study the consequences. We find that while a high cosmic ray flux does not affect the H II region by any significant amount relative to the heating and ionization due to the O stars, it plays an important role within the PDR and molecular regions. In atomic and molecular regions the main effects of cosmic rays are to heat and ionize the gas. The heating increases the gas temperature and the intensities of the infrared lines. Increased ionization can dissociate molecules, but also drive the chemistry faster due to the higher temperature and presence of ions. The combination of these effects leads to having an appreciable fraction of H$^0$ mixed in with the molecular gas, and to a considerable strengthening of the PDR emission lines. Figure 7 shows how the ([O I] λ146μm)/(Hβ) intensity ratio depends on the cosmic ray flux all the way from the Galactic background value up to the value for energy equipartition with the magnetic field.

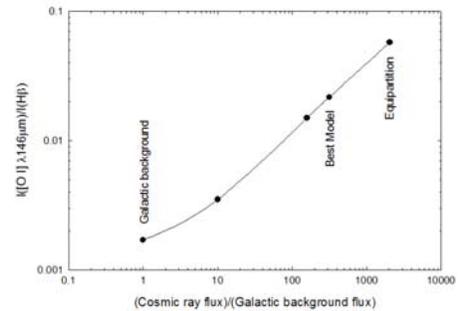

**Figure 7.** Dependence of the $I$([O I] λ146μm)/$I$(Hβ) intensity ratio on the cosmic ray flux.

Observations rule out the equipartition model. In the presence of the magnetic field, cosmic rays will produce synchrotron emission which will be strongest at longer radio wavelengths. We checked for evidence of synchrotron emission in the 330 MHz radio map measured for M17 by Subrahmanyan & Goss (1996). The observed ratio of the surface brightnesses at 1420 MHz (from Fig. 5) to that at 330 MHz is consistent with the ratio expected for just free-free and bound-free emission processes, and the morphological similarities between the maps at these two frequencies support the conclusion that most of the radio emission comes from $H^+$ zones. For the model in which the cosmic ray and magnetic energy densities are in equipartition, and assuming that the synchrotron emission comes from a region 1 pc deep along our line of sight, the expected synchrotron surface brightness at 330 MHz is equal to the measured peak value. Therefore, such a model would not be consistent with any radio continuum brightness contributed by the $H^+$ region itself.

However, for Model 2b with 316 times the Galactic background cosmic ray flux, the upper limit to the expected synchrotron surface brightness is 6.5 times fainter than the observed maximum. This surface brightness is about the same as the faintest measured values, and so is consistent with the 330 MHz observations. Model 2b also produces almost exactly the observed $I$([O I] $\lambda 146\mu m$)/$I$(H$\beta$) intensity ratio, so we adopt it as our best-fitting model, and list values calculated with it in Table 3.

These results provide empirical information about the behavior of cosmic rays in a magnetic field. Assuming that the ambient magnetic field started out at a typical ISM value of about 5-10 $\mu$G (Troland & Heiles 1986), $B$ has been increased by a factor 40-80. The cosmic ray density is about 300 times the Galactic background, so it appears that the cosmic ray density has responded to the compression of the magnetic field, but by an amount less than the energy equipartition case (in which case it would scale as $B^2$). The magnetic field must have been compressed quite rapidly by the pressure from starlight and stellar winds, and it is possible that some fraction of the cosmic rays leaked out in the process. Another possibility is that the cosmic rays have already lost a noticeable fraction of their energy through collisions with the gas in M17.

The upper panels of Figure 8 show Model 2b. The right-hand panel includes the OH density, to illustrate how OH and $H^0$ are mixed together. It is evident that if the 8-17 km s$^{-1}$ component stops at about 4 pc from the ionization front, and that the remaining gas is in the 20 km s$^{-1}$ component, the velocity structures of the $H^0$ and OH optical depth profiles (which are proportional to the column densities) would be similar to those seen in Figure 4.

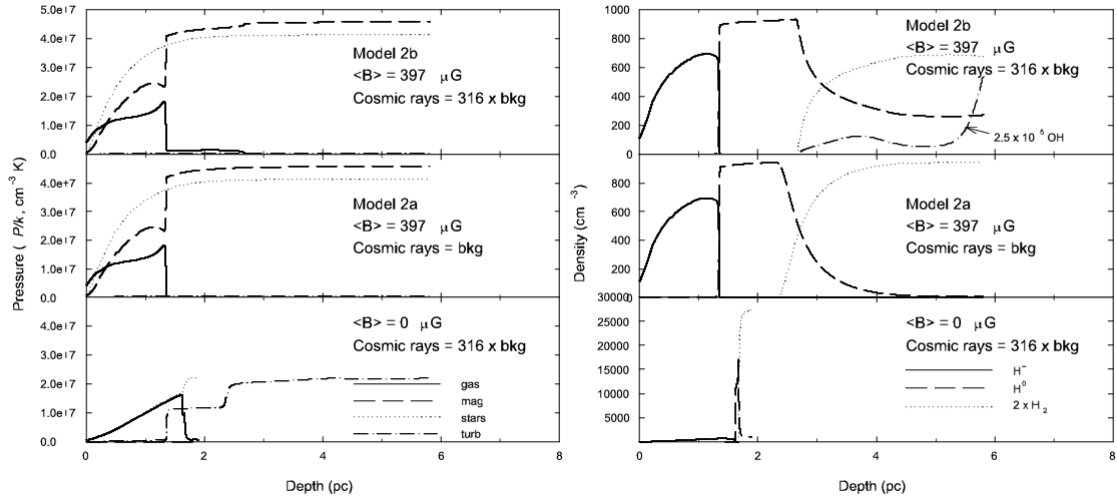

**Figure 8.** Pressure and density as a function of depth for Model 2b (top row), and for the same model but with cosmic rays set to the Galactic background level (Model 2a, middle row) and with the magnetic field set to 0 (bottom row). The format of the left-hand panels is similar to those in Figure 6, except that a dash-dot line (nearly coincident with the abscissa) shows $(P/k)_{turb}$, since this ratio is no longer set to be equal to $(P/k)_{mag}$. The format of the right-hand panels is also similar to those in Figure 6, except that the top-right panel includes an arbitrarily scaled OH density as a dash-dot line.

The middle and bottom rows in Figure 8 show the effects of decreasing the cosmic rays to the Galactic background value while still having the high magnetic field (this is the previously-described Model 2a), and alternatively of setting the magnetic field to zero while maintaining the high cosmic ray flux. It is clear that the strong magnetic field is the cause of the extended PDR in these models. While the high cosmic ray flux is also needed to match all of the observations, it does not by itself produce an extended PDR. The high density of cosmic rays probably is just a byproduct of the strong magnetic field in any case.

**8.5 The Sound Crossing Time**

Although we calculated a time-steady hydrostatic model as described in §6, we argued earlier in §2 that in fact the inner part of the H II region is expanding outwards and snowplowing into the 20 km s$^{-1}$ H$^0$ component. The sound crossing time through Model 2b from the illuminated face to the outer edge of the H II region is 90,000 years, while it is 650,000 years to a depth of 3pc (the point where the H$^0$ density has dropped by a factor of two), and about 5 million years to the outer edge of the whole model. However, the theoretical prediction is that a shock will be driven by the advancing ionization front, with the details depending on the exact circumstances in a rich and dynamic way (eg. García-Segura & Franco 1996; see also the discussion in §§6.5, 6.6 of AGN3). The observed line-of-sight velocity difference between the two H$^0$ components is 10 km s$^{-1}$. Allowing for the 45° viewing angle relative to the radial direction outwards from the stars, the actual velocity is 14 km s$^{-1}$, about Mach 3 in the PDR. If the shock wave currently has penetrated 4 pc into the cloud, as is estimated from the overlap between the OH and H$^0$ absorption profiles, the corresponding travel time is about 400,000 years. The present wave may be driven by the light from one or a very few hot O stars that turned on at this very recent time.

**8.6 The Gas-Phase Abundances in the PDR**

Table 3 shows that Model 2b gives a good match to the observed strength of [O I] λ145.5 μm, but overpredicts [C II] λ157.6 μm by a factor of 5 and underpredicts [O I] λ63μm and [Si II] λ34μm by a factor of 2. This could be due to simple abundance effects. Our models were calculated using the chemical composition described in §5.3. However, those abundances were intended as only an approximate description of what might be found in the gas (as opposed to dust) component of the PDR. We can compare the observed and predicted PDR line strengths from Table 3, and find the amount by which our assumed abundances need to be changed in order to fit the observations. In the PDR the emission lines of many different elements are sharing the load of cooling the gas, so a modest change in the abundance of any particular element will produce a directly proportional change in the strengths of the lines emitted by that element.

The two [O I] lines have an observed intensity ratio $I(\lambda 63\mu m)/I(\lambda 146\mu m) = 7.3$, as compared to an expected ratio in the range 9-24 (Abel et al. 2005; Kaufman et al. 1999). Meixner92 concluded that the $\lambda 63\mu m$ line is self-absorbed, which is verified by our model, so we discard this line from any abundance analysis.

Table 4 summarizes how the final abundances of O, C and Si were calculated. Column 3 lists the abundances assumed for Model 2b. Column 4 lists the ratio of the observed to the predicted line intensity, which is the factor by which we must correct the abundance in column 3 in order for Model 2b to exactly predict the observed line strength. Column 5 lists the abundances including this correction.

| Table 4 Abundance Determination | | | | |
|---|---|---|---|---|
| Line | Element | Abundance Assumed in Model 2b (X/H) | Observed/Predicted Line Strength | Final Derived Abundance (X/H) |
| [O I] λ145.5 μm | O | 3.4E-04 | 1.02 | 3.5E-04 |
| [C II] λ157.6 μm | C | 7.3E-05 | 0.19 | 1.4E-05 |
| [Si II] λ34.81 μm | Si | 4.1E-06 | 2.11 | 8.6E-06 |

After these adjustments to the abundances, we find that in the PDR, the C/O abundance ratio is 0.04. This ratio depends somewhat on the choice of the cosmic ray flux, ranging from C/O = 0.009 for a cosmic ray flux at the Galactic background level, to C/O = 0.07 for the cosmic ray flux in energy equipartition. All of these values are considerably lower than the ratio C/O = 1.7 found in the ionized gas (§5.3). Such a drop in the C/O ratio is expected if most of the carbon in the PDR has been depleted onto PAHs and/or grains, as would follow from the idea that grains grow in well shielded regions (Massa & Savage 1984; Cardelli & Clayton 1991).

We also calculated intensities for the observed $^{12}$CO J = 1–0, 2–1 and 3–2 lines. Model 2b predicts the three lines to have on average 0.07 times the observed intensities, after allowing for the lower C abundance found here. As can be seen in Figure 2, the measured CO emission at Position 1 is likely to include a contribution from background material that lies along the line of sight beyond the $H^+$ zone, so any model result lower than the observed CO strength is acceptable. In Model 2b, 90 percent of the CO emission would come from a background cloud that is not included in the model.

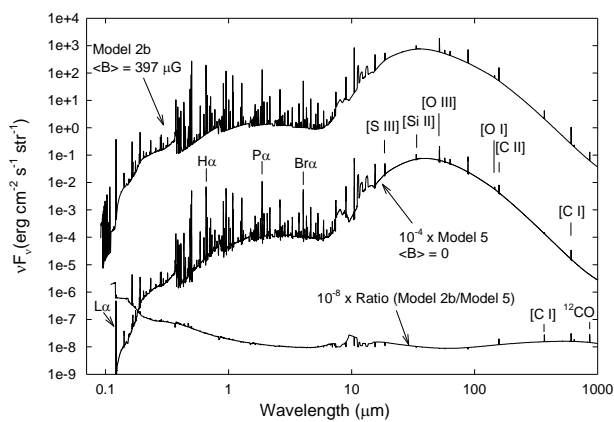

**Figure 9.** Predicted spectra from Model 2b and from Model 5 (which has no magnetic field), and their ratio. The models differ mainly in their continuum shapes at short wavelengths, in the PAH emission band near 10μm, and in the strengths of C and CO emission lines in the infrared.

### 8.7 The Predicted Spectrum

For comparison to future observations, Figure 9 shows the spectrum calculated for Model 2b over the wavelength range λλ0.1-2000 μm, and also the predicted spectrum of a model with no magnetic field (Model 5 from Figure 6).

Figure 10 shows how the local emissivity of several lines changes across the $H^+$ zone and PDR. This figure can be compared directly with the strip scans given by Meixner92 if we assume that the depth along our line of sight is constant.

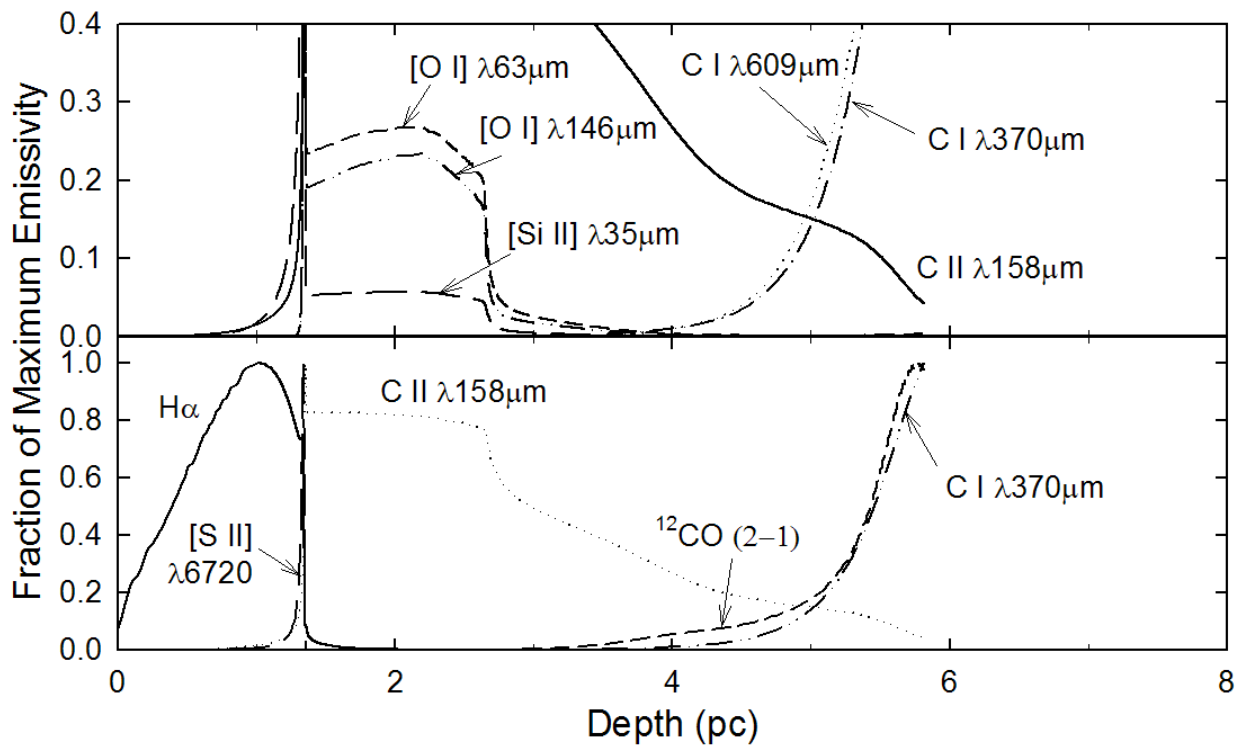

**Figure 10.** Predicted line emissivity per unit volume (normalized to their maximum values) as a function of depth in Model 2b.

### 9. Extended PDRs: Magnetic Fields *vs.* Clumpy Structure

A key feature of the SW bar in M17 is an extended PDR much thicker than is expected from simple models with smooth density distributions (as for example Model 5 in this paper, or the standard models by Tielens & Hollenbech 1985).

We have shown that the magnetic field that is also observed to be present in M17 will by itself lead to an extended PDR of the sort that is seen. As we have explained above, our final best model (Model 2b) incorporates a number of arbitrary choices and assumptions concerning such parameters as the incident radiation field, chemical abundances, magnetic field structure, and some details of the physics. We had to choose *something* in order to make the calculation. However, we have computed a far wider range of models than are shown in Table 3 or Figures 6 and 8, spanning a considerable range in almost all of these input parameters. We find that in all cases, once the magnetic field gets up to some critical value in the $<B> \sim 150$ μG range, magnetic pressure begins to control the PDR structure and a broad PDR is the inevitable result.

The model has been tuned to simultaneously match many observed parameters: the observed internal extinction, pressure from the x-ray gas, the projected thicknesses of the $H^+$ and $H^0$ (PDR) regions, the turbulent velocity, the $H^0$ and OH velocity structures, and the strength of the PDR lines. We note that the observed thicknesses of the $H^+$ and $H^0$ regions are upper limits, because we probably are not viewing the ionization front *exactly* edge-on. Also, the pressure from the x-ray gas is only an approximate calculation which depends on assuming the depth of the x-ray emitting region along the line of sight in order to calculate its total volume. The actual pressure acting on the illuminated face of the cloud could easily be different by a few tens of percent, plus the pressure balance is probably only approximate in any case. Nor do we accurately know the incident radiation field, because of uncertainties in the numbers of stars, as well as in their precise locations and surface temperatures. There is therefore a considerable range of choice for what would exactly be a best-fitting magnetically dominated model of the sort we propose here. We doubt that Model 2b is the only one that would fit all of the data. But the observed magnetic fields none-the-less imply that magnetic pressure must be taken into account.

Despite the choices described above, for the large majority of our simulations the magnetic field deep in the PDR came close to the peak observed value. This result does not depend on any details, only on our basic assumption that the overall geometry is roughly in hydrostatic equilibrium, so that the outward forces caused by gas and radiation pressure are balanced by the supporting pressure of the magnetic field. In fact, high magnetic fields are normally associated with regions of active star formation (Heiles & Crutcher 2005). The conclusion that M17 is in rough hydrostatic equilibrium leads to the simple picture that the formation of a cluster of stars pushed back the surrounding gas, enhancing both the magnetic fields and cosmic rays, until the magnetic field could resist the forces from the star cluster. The details of the dynamics and the cosmic ray transport are beyond the scope of this paper, but this general scenario is in agreement with all expectations.

Before these magnetic fields were measured, Stuzki et al. (1988) and Meixner92 developed a different explanation for the extended nature of M17's PDR. They argued that the dense clumps that they detected with their KAO observations and which had also been found at radio wavelengths are embedded in an extended, lower density medium. Meixner92 used the Tielens & Hollenbach (1985) code to compute a "homogeneous model" which is similar to our Model 5 ($<B>$ = 0). They argued that these conventional PDR models cannot explain the observed separation of the [O I], [C II] and CO emission peaks on the sky, or the large spatial extent of the [C II] emission measured by Stuzki et al. (1988). They also argued that the strong CO emission must come from regions of high density because of the high critical densities of these lines, that the high [Si II] $\lambda 35\mu m$ and [O I] $\lambda 146\mu m$ intensities imply that these lines come from regions of high density and temperature, while low ([O I] $\lambda 63\mu m$)/([C II] $\lambda 158\mu m$) intensity ratio shows that these lines come from different gas with lower densities and temperatures. To explain these features, they developed a "clumpy" model which essentially combined together three different PDR models, each calculated separately. These three regions represented dense ($n = 5 \times 10^5$ cm$^{-3}$) clumps, a moderate-density ($n = 3000$ cm$^{-3}$) "core" surrounding the clumps, and then a larger low-density ($n = 300$ cm$^{-3}$) molecular cloud surrounding the whole assemblage. By separately adjusting the parameters of these three regions, they fit the available observations about as well as does our model.

As an example of the detailed differences in interpretation from using the Meixner92 models as compared to our models, we note that from the ([C II] $\lambda 158$ $\mu m$)/([O I] $\lambda 146$ $\mu m$) intensity ratio at Position 1 and using the standard abundances in the Tielens & Hollenbach code, Meixner92 found that these lines are formed in gas with density $n_H = 2.3 \times 10^5$ cm$^{-3}$. With our model, using the final derived abundances listed in the right-hand column of Table 5 (which we argue are perfectly reasonable PDR abundances), we find that the same line ratio can be produced at a density $n_H \sim 900$ cm$^{-3}$.

Meixner92 showed that the observed [O I] $\lambda 63\mu m$ /$\lambda 146\mu m$ intensity ratio is not consistent with any optically thin model, and deduced that [O I] $\lambda 63\mu m$ is self absorbed. Although their three-component model did not include self-absorption effects, they argued that the lowest density component absorbs [O I] $\lambda 63\mu m$ emission originating in the higher density gas. Our models do find that the [OI] $\lambda 63\mu m$ line is optically thick. We account for the fact that an optically thick line is not emitted isotropically (Ferland et al. 1992). We find that the [O I] λ63μm line is strongly beamed, in the case of Model 2b having only 19 percent going outward with the remaining light being directed inward toward the ionizing stars. Meixner's observations presumably include emission from both the near and far side of the PDR shell, viewed at an angle as is sketched in Figure 2. Since the intensity of the beamed radiation will have a strong angular dependence, we should see a flux somewhere between the outward flux and sum of the inward and outward fluxes. The calculated fluxes listed in Table 3 are for the total inward + outward emission, but allowing for beaming the [O I] λ63μm /λ146μm ratio predicted by our model is actually in the range from 2.4 to 13.0, for the outward component or the total respectively, consistent with the observed ratio of 7.3.

It is abundantly clear from the observational data that there are clumps and condensations in the neutral and molecular gas, with size scales down to the angular resolution limit (Stutzki et al. 1988; Meixner92, BT99; BT01). We do not disagree with the approach used by Stuzki et al. (1988) and Meixner92 to model the clumpy nature of M17's PDR. We do not even attempt to model the observed fine-scale structure. Our Model 2a, which grossly underpredicts the strengths of the PDR lines, might represent just the low-density region postulated by Meixner92. Model 2b is presented as an alternative which, with a bit of juggling of abundances as described in §8.5, can in fact reproduce all of the observed line strengths. However, even in that case the observations show that additional, unmodeled condensations must be present. What we have attempted to show here is that now that we know that a strong magnetic field is present, it is important to include it in any models on any scale, because it has strong ramifications on the general structure of the PDR and molecular cloud.

The models presented in this study involve the compression of the PDR magnetic field by a combination of hot gas pressure and the momentum of ionizing radiation. This compression will naturally increase the field strength, and it is a consequence of the star formation process itself. However, high magnetic fields (relative to the interstellar average) are expected on other grounds in the M17 region, even in the absence of star formation effects. Only a rather small sample of molecular clouds has known magnetic field strengths via the Zeeman effect. These clouds appear to be in approximate balance between gravitational energy, magnetic energy, and the energy of internal motion (Myers & Goodman 1988a,b; Crutcher 1999). By implication, equipartition of this type arises naturally during the formation of self-gravitating molecular clouds, although the formation process is poorly understood. If the M17 SW cloud is in equipartition, then magnetic field strengths in the cloud should be in the range 500-1000 μG, values comparable to peak values measured by BT99 and BT01. In short, high magnetic fields in the M17 region may be a result of the molecular cloud formation process, not merely a result of subsequent star formation within the cloud. Of course, field strengths are unlikely to be uniform throughout the M17 SW cloud. They would presumably be highest in the central regions of the cloud and weaker near the periphery.

The M17 star cluster, in particular, must have formed in a part of the cloud where field strengths were considerably weaker than 500–1000 μG. First, the geometry of the region indicates that the star cluster formed in the periphery of M17 SW (Figures 1 and 2). Moreover, the morphology of $H^+$ region and molecular gas suggests that the star cluster has excavated a cavity in the molecular cloud, a cavity now filled with hot, X-ray emitting gas. The thermal pressure in the hot gas (§6.2 above) corresponds to a magnetic field of approximately 120 μG. Were the field at the position of the star cluster considerably higher than this value, then the magnetic field would have dominated the energetics and winds and light from the star cluster would have had little effect on the location of the gas. That is, no cavity could have formed. The fact that the thermal pressures are about the same in the hot gas and the HII region suggests that thermal gas pressure dominates over magnetic pressure, at least in the ionized-neutral gas interface region. For this circumstance to be true, the ambient field near the star cluster must have been less than about 120 μG. In fact the measured line-of-sight fields in the regions around the cluster are only of order 50 μG. We conclude that the star cluster formed in the outer regions of the M17 SW cloud where the field was considerably weaker than the equipartition field in the central regions of the cloud. In such an environment, the combination of hot gas pressure and momentum in starlight was able to push back the gas, strengthening the local field, and producing the observed geometry.

## 10. Conclusions

One has only to look at the M17 images presented here and in many other papers to see that we are dealing with a complex structure in which many things are going on. We are concerned here with understanding the major forces that shape the overall structure, not the details that cause the many condensations and other sub-structures. We propose that the radiation and wind from the star cluster is what has determined the present structure of M17. The same process is happening in the Orion Nebula (Baldwin et al. 1991; O'Dell 2001). The observed magnetic field strength shows that magnetic pressure is an important factor in controlling the overall structure of the PDR (and presumably also of the molecular gas). Clumps such as those shown in direct images (Meixner92) represent changes in the gas pressure that are insignificant compared to the pressure in the surrounding magnetic field. The observed condensations are only fluff floating on top of the sea that is the magnetic field.

We have shown that above a fairly low threshold in the magnetic field (roughly $<B> = 150$ µG for the incident radiation fields present in M17), the magnetic pressure takes over from gas pressure in supporting the PDR and molecular cloud. Ambient magnetic fields of far lower strength (a few µG) are compressed by the external pressure on the cloud until pressure balance is achieved (§7.2). In the case of M17, some external pressure is exerted on the illuminated face of the $H^+$ zone by the hot x-ray emitting gas that surrounds the ionizing stars, but it is the absorption of the momentum in the cluster's starlight that leads to the large pressures seen in the magnetic field. In all of the models shown here, gas pressure still dominates in the H II region, but in fact we did run some extreme cases in which magnetic pressure dominated this zone as well.

We then fit a fairly detailed model to Position 1 in M17. The major parameter for which there is not at least a reasonable estimate from observations or basic physics is the cosmic ray flux, which we had to set to 300 times the Galactic background level in order to match the observed PDR line strengths and the observed mixing of $H^0$ and OH. We did not attempt to model the observed clumpy structure, and as Meixner92 and others have shown, such structure will also affect the PDR line strengths. Changing the proportions of O *vs.* B stars in the incident spectrum would also have some effect, due to the resulting changes in the shape of the incident continuum. Due to the many parameters involved, we do not claim that our model is unique. However, our model does demonstrate that it is indeed possible to match a wide variety of observational constraints with a magnetically supported model.

The pressure in the observed magnetic field just balances the outward forces. In our view, the sequence of events was: the molecular cloud was there; the stars turned on and put pressure on the surrounding gas through a combination of photons and a stellar wind; the surrounding gas was pushed back and compressed, compressing the magnetic field with it, until the magnetic pressure balanced the pressure from the stars. The observed geometry and high magnetic field are then a natural consequence of the formation of a star cluster within a molecular cloud, and therefore may be a feature of many such systems.

If all other things are equal, the thicknesses of edge-on PDRs should be an indicator of the magnetic pressure. The thickness can be measured in a much wider variety of situations than it is possible to directly measure the magnetic field strength. For example, we are currently measuring the PDR thickness in NGC 3603 and 30 Doradus. A long term goal of our work is to learn how to better interpret the spectra of giant starbursts seen at high redshifts. 30 Dor is the largest and most luminous Giant H II Region in the Local Group of galaxies, and so is of particular interest as a nearby example of similar objects. 30 Dor has a very complex structure, but an intriguing result is the report by Poglitsch et al. (1995) that the PDR gas extends very far into and is commingled with the molecular gas. Can a magnetic field be at work? If so then it provides the fundamental support that accounts for the observed structures. Magnetic pressure support, together with the assumption of hydrostatic equilibrium, is a much simpler situation than has previously been pictured, and would lead to better insight into the message in the spectrum.

We are grateful to Will Henney, John Mathis, Mark Voit, Robin Williams and Ellen Zweibel for helpful discussions, and to the anonymous referee for very valuable suggestions. E.W.P and J.A.B. acknowledge financial support from NSF through grant AST-0305833 and from NASA through HST grant GO09736.02-A. M.M.H. gratefully acknowledges support from the National Science Foundation under grant AST-0094050 to the University of Cincinnati. GJF thanks NSF for support through AST -0607028 and NASA for support through NNG05GG04G. T.H.T. acknowledges financial support from NSF grant 03-07642.

**Appendix. The Stellar Lyman Lines and the 21 cm Spin Temperature**

We would have preferred to have used an observed $H^0$ column density as our stopping criterion. However, this was not possible because the radio observations only determine the integral

$$\tau(21\,\mathrm{cm}) = \int \frac{n(H^0)}{T_{spin}} dl \quad (A1)$$

where $T_{spin}$ is the 21 cm spin temperature and the integral is over the cloud. It is well known that scattered Ly$\alpha$ can affect $T_{spin}$ by pumping the hyperfine levels of H via the 2p state (Field 1959). Our simulations include this effect since we fully consider the internal structure of all atoms and ions of the H-like (Ferguson & Ferland 1997) and He-like isoelectronic sequences. Excited levels of $H^0$ can be populated following recombination, collisional excitation (mainly by cosmic rays) or by fluorescence induced by the continuum. Most such excitations will eventually produce Ly$\alpha$ photons (§4.2 of AGN3), which can then affect the 21 cm spin temperature. Further details are given in Shaw, Ferland & Srianand (2006, in preparation).

Of the processes that create Ly$\alpha$, the most important and uncertain is the stellar continuum. Depending on the precise wind and thermal structure of the stellar atmosphere, the Lyman lines can be either in emission or absorption. The COSTAR atmospheres we use for the hottest stars do not include Lyman lines at all. The continuum at the wavelengths of the cores of the Lyman lines can pump excited levels of $H^0$, producing Ly$\alpha$, which scatters to raise $T_{spin}$.

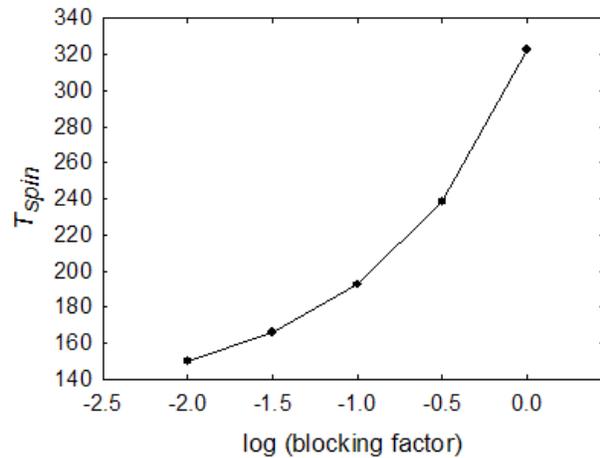

In the calculations presented here, we assumed that the stars in the M17 cluster do not radiate in the Lyman lines. This is equivalent to assuming that the Lyman lines are strongly in absorption. The published continua have Lyman lines weakly in absorption, with the center of the line depressed by roughly 10 percent relative to the neighboring continuum. But these models do not include the effects of winds, which could dramatically alter the Lyman line profiles.

Figure 11 shows the effect on $T_{spin}$ of varying the assumed Lyman-line absorption for a model similar to Model 2 (but with an incident flux two times higher). The x-axis is the log of a "blocking factor" which multiplies the actual continuum at the wavelengths of the Lyman lines. Unity means that the continuum has the value in the published model. The 21 cm spin temperature changes by a factor of two. In model 2b, the mean spin temperature with the Lyman lines fully blocked is 195 K, while the true $H^0$-weighted kinetic temperature is 192 K.

**Figure 11.** The effect on the 21 cm spin temperature $T_{spin}$ of varying the blocking factor for the Lyman absorption lines in the stellar atmospheres. Log (blocking factor) = 0 corresponds to no Lyman line absorption, while smaller values correspond to increasingly stronger absorption lines.

The ISM prevents direct observation of Lyman lines in hot stars. This lack of observations means that there are few published studies of the properties of Lyman lines in early stars. However, it is still very important to calculate their strengths because of their major influence on conditions in the interstellar medium.